\documentclass[a4paper,12pt]{article}
\usepackage{epsfig}
\usepackage{amssymb}
\usepackage{amsfonts}
\usepackage{amsmath}
\usepackage{euscript}
\usepackage{verbatim}
\usepackage{latexsym}
\usepackage{graphicx}
\usepackage{caption, color}
\usepackage{float}
\usepackage{cite, hyperref}
\usepackage{subcaption}

\usepackage[all]{xy}

\hypersetup{ linktoc=all,
    colorlinks, linkcolor={palatinateblue},
    citecolor={palatinateblue}, urlcolor={amaranth}
}

\graphicspath{{Images/}}

\definecolor{rosy}{RGB}{230,235,252}
\definecolor{myframetitle}{RGB}{90,89,170}
\definecolor{myblocktitle}{RGB}{140,185,249}
\definecolor{mytitle}{RGB}{10,80,26}

\definecolor{darkgreen}{RGB}{27,130,45}
\definecolor{darkblue}{rgb}{0,0,0.3}
\definecolor{darkred}{rgb}{0.7,0,0}

\definecolor{light gray}{RGB}{220,220,220}
\definecolor{dark purple}{RGB}{108,0,217}
\definecolor{pink}{RGB}{190,20,100}
\definecolor{orang}{RGB}{193,63,0}
\definecolor{green}{RGB}{11,98,17}
\definecolor{darkpink}{RGB}{153,0,76}
\definecolor{bluegreen}{RGB}{0,102,102}
\definecolor{greenlagan}{RGB}{0,102,0}
\definecolor{redgreen}{RGB}{102,102,0}
\definecolor{Redgreen}{RGB}{153,76,0}
\definecolor{vividviolet}{rgb}{0.62, 0.0, 1.0}
\definecolor{amaranth}{rgb}{0.9, 0.17, 0.31}
\definecolor{palatinateblue}{rgb}{0.15, 0.23, 0.89}
\definecolor{brightpink}{rgb}{1.0, 0.0, 0.5}
\definecolor{cornflowerblue}{rgb}{0.39, 0.58, 0.93}
\definecolor{deepcarminepink}{rgb}{0.94, 0.19, 0.22}
\definecolor{radicalred}{rgb}{1.0, 0.21, 0.37}

\usepackage{listings}

\newif\ifdtup

\catcode`\@=11

\@addtoreset{equation}{section}

\def\@normalsize{\@setsize\normalsize{15pt}\xiipt\@xiipt
\abovedisplayskip 14pt plus3pt minus3pt%
\belowdisplayskip \abovedisplayskip
\abovedisplayshortskip \z@ plus3pt%
\belowdisplayshortskip 7pt plus3.5pt minus0pt}

\def\small{\@setsize\small{13.6pt}\xipt\@xipt
\abovedisplayskip 13pt plus3pt minus3pt%
\belowdisplayskip \abovedisplayskip
\abovedisplayshortskip \z@ plus3pt%
\belowdisplayshortskip 7pt plus3.5pt minus0pt
\def\@listi{\parsep 4.5pt plus 2pt minus 1pt
     \itemsep \parsep
     \topsep 9pt plus 3pt minus 3pt}}

\relax

\catcode`@=12

\catcode`\@=11

\def\section{\@startsection{section}{1}{\z@}{3.5ex plus 1ex minus
   .2ex}{2.3ex plus .2ex}{\large\bf}}

\def\SymBoxes#1#2#3#4{\newdimen\un@t \un@t#3%
\raisebox{#1}{\rule{#2\un@t}{#4}\hskip-#2\un@t
\@tempdimb\un@t \advance\@tempdimb by-#4\@tempcntb#2\relax%
\@whilenum{\@tempcntb>0}\do{
\rule{#4}{\un@t}\hskip\@tempdimb \advance\@tempcntb by\m@ne}%
\hskip-#2\un@t \rule[\un@t]{#2\un@t}{#4}%
\rule[\un@t]{#4}{#4}\hskip-#4
\rule{#4}{\un@t}}\hskip-#4}                

\begin{document}

\newcommand{\beq}{\begin{equation}}
\newcommand{\eeq}{\end{equation}}
\newcommand{\bea}{\begin{eqnarray}}
\newcommand{\eea}{\end{eqnarray}}
\newcommand{\beas}{\begin{eqnarray*}}
\newcommand{\eeas}{\end{eqnarray*}}
\newcommand{\defi}{\stackrel{\rm def}{=}}
\newcommand{\non}{\nonumber}
\newcommand{\bquo}{\begin{quote}}
\newcommand{\enqu}{\end{quote}}
\renewcommand{\(}{\begin{equation}}
\renewcommand{\)}{\end{equation}}
\def \eqn#1#2{\begin{equation}#2\label{#1}\end{equation}}

\def\e{\epsilon}
\def\IZ{{\mathbb Z}}
\def\IR{{\mathbb R}}
\def\IC{{\mathbb C}}
\def\IQ{{\mathbb Q}}
\def\de{\partial}
\def\Tr{ \hbox{\rm Tr}}
\def\H{ \hbox{\rm H}}
\def\HE{ \hbox{$\rm H^{even}$}}
\def\HO{ \hbox{$\rm H^{odd}$}}
\def\K{ \hbox{\rm K}}
\def\Im{ \hbox{\rm Im}}
\def\Ker{ \hbox{\rm Ker}}
\def\const{\hbox {\rm const.}}
\def\o{\over}
\def\im{\hbox{\rm Im}}
\def\re{\hbox{\rm Re}}
\def\bra{\langle}\def\ket{\rangle}
\def\Arg{\hbox {\rm Arg}}
\def\Re{\hbox {\rm Re}}
\def\Im{\hbox {\rm Im}}
\def\exo{\hbox {\rm exp}}
\def\diag{\hbox{\rm diag}}
\def\longvert{{\rule[-2mm]{0.1mm}{7mm}}\,}
\def\a{\alpha}
\def\dag{{}^{\dagger}}
\def\tq{{\widetilde q}}
\def\p{{}^{\prime}}
\def\W{W}
\def\N{{\cal N}}
\def\hsp{,\hspace{.7cm}}

\def\br{\nonumber}
\def\IZ{{\mathbb Z}}
\def\IR{{\mathbb R}}
\def\IC{{\mathbb C}}
\def\IQ{{\mathbb Q}}
\def\IP{{\mathbb P}}
\def \eqn#1#2{\begin{equation}#2\label{#1}\end{equation}}

\newcommand{\C}{\ensuremath{\mathbb C}}
\newcommand{\Z}{\ensuremath{\mathbb Z}}
\newcommand{\R}{\ensuremath{\mathbb R}}
\newcommand{\rp}{\ensuremath{\mathbb {RP}}}
\newcommand{\cp}{\ensuremath{\mathbb {CP}}}
\newcommand{\vac}{\ensuremath{|0\rangle}}
\newcommand{\vact}{\ensuremath{|00\rangle}                    }
\newcommand{\oc}{\ensuremath{\overline{c}}}
\newcommand{\psizero}{\psi_{0}}
\newcommand{\phizero}{\phi_{0}}
\newcommand{\hzero}{h_{0}}
\newcommand{\psiin}{\psi_{\rh}}
\newcommand{\phiin}{\phi_{\rh}}
\newcommand{\hin}{h_{\rh}}
\newcommand{\rh}{r_{h}}
\newcommand{\rb}{r_{b}}
\newcommand{\psibnd}{\psi_{0}^{b}}
\newcommand{\psibndp}{\psi_{1}^{b}}
\newcommand{\phibnd}{\phi_{0}^{b}}
\newcommand{\phibndp}{\phi_{1}^{b}}
\newcommand{\gbnd}{g_{0}^{b}}
\newcommand{\hbnd}{h_{0}^{b}}
\newcommand{\zh}{z_{h}}
\newcommand{\zb}{z_{b}}
\newcommand{\man}{\mathcal{M}}
\newcommand{\hbr}{\bar{h}}
\newcommand{\tbr}{\bar{t}}

\newcommand\tcr{\textcolor{red}}
\newcommand\tcb{\textcolor{blue}}
\newcommand\tcg{\textcolor{green}}

\newcommand\snote[1]{\textcolor{red}{\bf [Sh:\,#1]}}

\begin{titlepage}
\begin{flushright}
CHEP XXXXX 
\end{flushright}
\bigskip
\def\thefootnote{\fnsymbol{footnote}}

\begin{center}
{\large{\bf Young Black Holes Have Smooth Horizons:  \\ 
A Swampland Argument }}
\end{center}

\bigskip
\begin{center}
Chethan Krishnan$^a$\footnote{\texttt{chethan.krishnan@gmail.com}}, \ \ Ranjini Mondol$^a$\footnote{\texttt{ranjinim@iisc.ac.in}}, 
\vspace{0.1in}

\end{center}

\renewcommand{\thefootnote}{\arabic{footnote}}

\begin{center}

$^a$ {Center for High Energy Physics,\\
Indian Institute of Science, Bangalore 560012, India}\\


\end{center}

\noindent
\begin{center} {\bf Abstract} \end{center} 
It has been suggested that black hole microstates in string theory may have ``no interiors". The arguments of arXiv:2312.14108 suggest that even if this were the case, conventional bulk EFT should be valid at the horizon for black holes that are not too old. Here, we explore the validity of bulk EFT for young black holes from a different perspective, by considering the gravitational collapse of scalar moduli. This builds on arXiv:2003.05488, which studied the numerical collapse of a scalar field in a Choptuik-like set up. Evidence was presented there for two observations -- (a) For every scalar profile that results in collapse, the scalar field undergoes trans-Planckian variation, (b) Every trans-Planckian scalar motion is hidden behind an apparent horizon. The results of arXiv:2003.05488 were mostly obtained in regimes close to critical collapse, where scale-separation was difficult to achieve. In this paper, we study the same system in super-critical regimes. This allows us to study the localization properties of scalar variations inside the apparent horizon by separating the horizon and singularity scales. We find strong evidence that the $\mathcal{O}(1)$ scalar variation is localized around the curvature divergence, and is hierarchically well-separated from the apparent horizon. Because $\mathcal{O}(1)$ scalar movement is associated to the breakdown of bulk EFT according to the swampland distance conjecture, this is an indication that bulk EFT is well-behaved in the interiors of young black holes. Our results suggest the perspective that cosmic censorship is a mechanism for preserving bulk EFT between the horizon and the singularity.

\vspace{1.6 cm}
\vfill

\end{titlepage}

\setcounter{footnote}{0}

\tableofcontents 

\section{Introduction}

The principle of equivalence says that the event horizon is a ``smooth" region in a black hole spacetime. This means that one can write down coordinates which are locally inertial through the horizon, Kruskal metric being an illustration of this general idea.

The principle holds in general relativity, or more broadly, in bulk Effective Field Theory (EFT). 
Because the smoothness of the horizon is a statement about bulk EFT, it is not {\em a-priori} necessary \cite{Burman} that there is an analogous statement in the UV-complete bulk description\footnote{The distinction between bulk EFT and the bulk UV-complete description when discussing the black hole interior  is often quite murky in the litreature. Our discussion is closest in spirit to the Introduction of \cite{Witten}, see also \cite{Burman}. We will use the phrases ``UV complete bulk description" and ``non-perturbative string theory" interchangeably -- finite Planck scale in the bulk implies finite-$N$. There exists the possibility that non-perturbative string theory only has a holographic description and not a bulk description, but we will ignore this possibility in this paper.}. However, since we expect that bulk EFT should emerge as a limit of the UV-complete description, it is a reasonable assumption that smoothness of the horizon should also emerge in such a limit (whatever that limit may eventually turn out to be).

There have been various suggestions across decades that black holes may not have interiors \cite{tHooft, Susskind, MathurFuzz}. The original source of one version of the idea is 't Hooft \cite{tHooft}. In present day string theory, a realization of such an idea exists -- this is the fuzzball program of Mathur and collaborators, which aims to construct microstates of black holes in the bulk. The geometric microstates constructed so far in supergravity have the feature that they do not have an interior\footnote{Since these are supergravity solutions, their precise status in the quantum theory (with finite-$N$) is unclear. Our perspective on them is simply that they are an existence proof that no-hair theorems can be evaded.}. The precise relationship between these supergravity microstates and the smooth horizon as experienced by a probe field (or an infaller) has not been clarified, but there have been suggestions \cite{Fuzzcomplementarity}. 

In \cite{Burman} (see also \cite{Krishnan}) the question of the emergence of the interior from a quantum horizon was addressed more broadly\footnote{Closely related previous and subsequent work is in \cite{DasSynthetic, DasFuzz, Das, DasBrickwall, SB}.}. Given that it is not known how to construct truly quantum microstates in the bulk, a Planckian stretched horizon was simply postulated as a model for a ``quantum" horizon for large black holes in AdS/CFT. It was argued that highly excited typical pure states (at an energy sliver at the mass of the black hole) on the stretched horizon vacuum are models of black hole microstates in the CFT. The correlation functions in this set up are therefore models for correlators in the UV-complete setting. Remarkably, it was found that these correlators are  indistinguishable (upto corrections exponentially suppressed in the black hole entropy) from the smooth horizon (Hartle-Hawking) correlator all the way to the Page time. This gives an understanding of the smoothness of the horizon in the classical limit as an emergent phenomenon, in the large-$N$ (or Page time $\rightarrow \infty$) limit. The calculation also gave an understanding of why we could get away without knowing the details of the UV complete description -- up to the Page time, these details do not play a role in determining the correlators. 

These ideas are of immediate relevance for the black hole information paradox \cite{Hawking, Page, Mathur, AMPS}, which can be viewed as a question about the breakdown of bulk EFT at sub-Planckian scales. While Hawking's original argument is sometimes viewed as not sophisticated enough to actually be called a contradiction in bulk effective field theory, it has been sharpened by various authors since \cite{Page, Mathur, AMPS}. The version of the paradox suggested in \cite{AMPS} implies that the absence of the interior arises as a sharp contradiction only for old black holes, older than their Page time. Various authors have claimed that the horizon is smooth even beyond the Page time, but a clear consensus (and in our opinion, a fully satisfying counter-argument) is still lacking.

While the situation {\em after} the Page time is murky, the results of \cite{Burman} are a strong indication that interior-less black hole microstates can reproduce the smooth horizon expectations from bulk EFT {\em before} the Page time. It should be emphasized that this result is (at least superficially) surprising -- we started in a set up without a manifest interior, and yet ended up finding the smooth horizon correlator, up to exponential corrections.  Therefore it is worth looking for other pieces of evidence that suggest that the horizons of not-too-old black holes are smooth in bulk EFT. Considering the fact that the fuzzball proposal claims that the UV complete bulk description of a microstate has no interior, the smoothness of the horizon is not a foregone conclusion. Indeed, the original suggestion in \cite{Mathur} was simply that bulk EFT must break down at the horizon scale, if the unitarity of the Page curve is to be preserved. There was no specific statement about young vs. old black holes\footnote{It is also worth pointing out that the AMPS argument {\em assumes} the smoothness of the horizon at early times in order to derive a contradiction with smoothness after the Page time.}.  


Of course, to qualify as a {\em new} piece of evidence for the smoothness of young horizons, we need to go beyond bulk EFT itself and look for constraints on bulk EFT that arise from the existence of the UV-completion.
The string swampland \cite{Vafa} is a collection of ideas that imposes restrictions on bulk EFT from UV completion considerations in string theory. It is a natural place for us to look for an argument of the kind we seek. The goal of this paper is to note that suggestive swampland-based observations do indeed exist, which indicate that the horizons of young black holes are smooth. 

Our key strategy will be to consider the gravitational collapse of a modulus scalar field. The advantage of considering a modulus, is that there are non-trivial and highly tested conjectures on the behavior of massless scalar field ranges in string theory. In particular, it is believed that if a scalar field value moves by an $\mathcal{O}(1)$ field range in Planck units during its evolution, then the spacetime region in which this happens cannot all be described in terms of a single low energy EFT \cite{Ooguri}. We would like to exploit this as a means of studying the breakdown of bulk EFT in spacetime regions containing black holes.

The massless scalar-gravity system is not an easy one to study, even if one assumes spherical symmetry. The reason is that one has to deal with partial differential equations (PDE). Most of the early work on gravitational collapse, eg. \cite{Dutt, Oppenheimer, Vaidya}, therefore assumed simplified models for stress tensors. While these can be instructive, their regime of reliability is often uncertain \cite{CK}. We would like to study collapse in a setting where we have a good understanding of the trustworthiness of the ingredients. We wish to keep track of the field range moved by the scalar during its evolution and collapse, and use it as a tracker for the breakdown of bulk EFT.  This means that we will have to confront the scalar-gravity system and its PDEs directly. 

The collapse of a massless scalar coupled to the gravitational field, was famously first studied by Choptuik \cite{Choptuik}. Choptuik discovered that as one varies initial conditions for the evolution (ie., the initial profile of the scalar field), the evolution switches over from disbursal to collapse. At the transition, there is a naked singularity, and near the transition there is critical behavior with precise critical exponents\footnote{This is a conceptually significant result in relativity that seems to be only accessible via numerical methods.}. In \cite{CK} we studied this system from the perspective of the string swampland. The goal was to test the distance conjecture in a dynamical setting and see that the $\mathcal{O}(1)$ motion of a scalar field is indeed associated to gravitational collapse as expected on general grounds \cite{Banks}. In every case that was investigated in \cite{CK}, it was found that $\mathcal{O}(1)$ scalar motion results in gravitational collapse as characterized by divergence in curvatures at late times\footnote{Note that it is not necessary that gravitational collapse only happens due to scalar motion. One can in principle cause collapse by colliding gravitational waves. But in the parametric families of solutions one studies by varying scalar profiles, the $\mathcal{O}(1)$ motion and collapse were observed to be precisely correlated \cite{CK}.}. Remarkably, it was further noted that in every such case, the $\mathcal{O}(1)$ motion was hidden behind an apparent horizon. 


In this paper, we will revisit this system. Most of the work in \cite{CK} was done close to criticality. This is the case that was numerically most accessible. But it also suffers from the drawback that the horizon and singularity scales essentially coincide. Since our goal in the present paper is to study the behavior of the scalar field range while paying special attention to differentiate the horizon and the singularity\footnote{Our goal is after all to study the well-definedness of the black hole ``interior", which is defined as the region between the horizon and the singularity.}, we will study collapse in the deeply super-critical regime. We find that this regime is numerically more challenging because the total evolution time to collapse can increase significantly -- this is the price one must pay, to make a clear statement about the scale of breakdown of bulk EFT inside collapsing black holes. In our examples, we work with supercritical regimes where the horizon scale is a few times to about 1-2 orders of magnitude bigger than the other relevant scales in the problem. In future work it will be good to improve on this front so that the curvature divergence is strictly orders of magnitude separated from the horizon scale. Even though our scale separation is not as clean as we would have liked, from our experience with the curves, we feel confident that the result is robust.

The question that we are interested in will be the following -- how is the scalar field motion localized in relation to the locations of the apparent horizon and the singularity? If the $\mathcal{O}(1)$ field range associated to the collapse is localized near the horizon (or is spread over the region between the horizon and the singularity), there would be a legitimate case to be made, that bulk EFT has broken down in the black hole interior. We find instead that this is never the case in any of the (large number of) cases that we have investigated. The $\mathcal{O}(1)$ scalar motion is invariably localized at or near the location of the singularity, and far from the location of the apparent horizon\footnote{By ``far" here, we mean hierarchically larger than the distance scale between the scalar growth and the curvature growth.}. Because the results are numerical and the swampland distance bound is conjectural, this cannot be called a proof, but it is strong evidence for the validity of bulk EFT in the black hole interior.

In the next section we will outline some aspects of the scalar-gravity system that we solve. The system we study and the numerical methods we employ are largely identical to those in \cite{CK}. So we will only discuss the points that are worthy of emphasis, a reader who is interested in the details should consult \cite{CK}. Our focus here will be on the key results, which are presented in Section \ref{Results}. Even here, much of the technical details that are less significant for the main line of development are relegated to the appendices. A concluding section recaps the main message of the paper.

\section{Supercritical Collapse of the Scalar Field}

The system we will numerically solve (see eg. \cite{Frolov, Guo, Guo2, Joshi}) is given by the action \cite{CK}
\begin{align}
    S = \frac{1}{8\pi G}\int d^{4}x\sqrt{-g}\Big(\frac{R}{2}-\frac{1}{2}\partial_{\mu}\phi\partial^{\mu}\phi\Big)
\end{align}
We have extracted an overall $8 \pi G$ for convenience. This means that $\phi/ \sqrt{8 \pi G}$ is the conventionally normalized scalar. The system has a classical scale invariance, which we will fix by setting $G=1$ and this defines our unit. 
The metric is taken as in \cite{CK}
\begin{align}
    ds^{2} = e^{-2\sigma}(-dt^{2}+dx^{2}) + r^{2}d\Omega^{2}
\end{align}
where $r$ and $\sigma$ are both functions of both $t$ and $x$. The equations of motion can be found in \cite{CK}. We will only write the Einstein constraints here because we have had to be somewhat more careful with them in this paper, compared to \cite{CK}: 
\begin{subequations}\label{Constraint}
    \begin{align}
        f_1\equiv \frac{\partial^{2}r}{\partial t \partial x}+\frac{\partial r}{\partial x}\frac{\partial \sigma}{\partial t}+\frac{\partial r}{\partial t}\frac{\partial \sigma}{\partial x} + 4\pi r \frac{\partial \phi}{\partial t}\frac{\partial \phi}{\partial x} = 0 \label{Constraint-1}
    \end{align}
    \begin{align}
        f_2 \equiv \frac{\partial^{2}r}{\partial t^{2}} +  \frac{\partial^{2}r}{\partial x^{2}} + 2\frac{\partial r}{\partial t}\frac{\partial \sigma}{\partial t} + 2\frac{\partial r}{\partial x}\frac{\partial \sigma}{\partial x} + 4\pi r\Big(\Big(\frac{\partial \phi}{\partial t}\Big)^{2} + \Big(\frac{\partial \phi}{\partial x}\Big)^{2}\Big) = 0 \label{Constraint-2}
    \end{align}
\end{subequations}
At each step of the evolution, we check that the constraints are small enough. The last step of the evolution is the one where one of these constraints goes above some pre-defined cut-off that we set. This happens in our collapsing geometries because the curvature scalar and the modulus scalar both start diverging. We present the plots of these constraints for a few time steps at the final stages of collapse for a Canonical Example (introduced in the next section), in Appendix \ref{Appendix-Constraints}. 

As in \cite{CK}, the Misner-Sharp mass $m$ is used as an auxiliary variable \cite{Frolov, Guo, Joshi, Csizmadia}. It is defined below. It is also useful for defining the location of the apparent horizon as the zero of the following quantity:
\begin{align}
    g^{\mu\nu}r_{,\mu}r_{\,\nu} = e^{2\sigma}\Big(-\Big(\frac{\partial r}{\partial t}^{2}\Big)+\Big(\frac{\partial r}{\partial x}^{2}\Big)\Big) \equiv 1-\frac{2 m}{r}
\end{align}
With this definition, the number of variables increases to four, but the stability of the integration is improved. By algebraic manipulations, the equations of motion and the constraints \eqref{Constraint} can be reduced to a set of equations (see \cite{CK}) that are better suited for finite difference integration. We will not repeat the equations here, nor discuss the methods for integration and the methods for setting initial/boundary data. They are identical to the ones discussed in \cite{CK} and the interested reader can consult there. We will however make one comment. Our interval of integration is from the origin $x=0$ to some positive $x_f$. Due to the causal structure of the hyperbolic PDEs that are being integrated, once we set up a consistent set of initial data (and boundary data at the origin) to start the integration, we can define our boundary conditions at $x_f$ by a process called ``extrapolation" \cite{Frolov, Guo, Guo2, Joshi, CK}. We need to be careful that the time evolution is only until the regime where the errors from the right boundary have not propagated to the interior region that we are interested in. We check that this is the case while making our plots and only consider regions of $x$ that are robust in this sense, in our plots. In this regime the evolution is independent of the data at $x_f$ \cite{Frolov, Guo, Guo2, Joshi, CK}.

We will mostly discuss the Gaussian profile in this paper. But we have also tested our code with the other profiles discussed in \cite{CK} and the conclusions are similar. We will refer to an evolution by specifying the values $A$, $x_0$ and $\delta$ that go into the (Gaussian) profile:
\begin{align}
    \frac{\phi(0,x)}{\sqrt{8\pi}} = A \exp{\Big(-\frac{(x-x_{0})^{2}}{\delta^{2}}\Big)} \label{Gauss}
\end{align}

\section{Results}\label{Results}

We will describe our main result using a sample evolution, see figure \ref{fig:phi_K_Vs_x_A_p_05}. We will refer to it as the Canonical Example. Some other examples are presented in Appendices\footnote{We have noticed that irrespective of the initial profiles that we choose, there are a few universality classes of final shapes that our curves evolve to. It may be interesting to make this precise and more exhaustive, but that is beyond our goals in this paper.}.

The key observations should be clear from the plots. The main message is that the growth of the modulus and the Kretschmann scalar divergence essentially coincide, while the apparent horizon is (hierarchically) far from both. This is a universal observation in every collapsing curve that we have seen, with the expected caveat that when we are close to criticality (ie., naked singularity), all three can overlap due to the lack of scale separation.

The plots here are in the $x$-coordinate, but we have checked the above statement also with more physically meaningful notions of separation. These include the radial variable\footnote{Note that it measures the size of the sphere.} $r$ and also the invariant distance 
\bea
s=\int dx \exp(- \sigma). \label{invdis}
\eea
See Appendix \ref{PhysDis} for a discussion on this. The modulus growth/Kretschmann divergence is well separated from the apparent horizon in terms of these physical distances as well, as long as we are not too close to critical collapse.

Evolving the system becomes harder as time steps get closer to the singularity,  so in some cases we have not been able to explicitly check that the divergence of the modulus gets to the $\mathcal{O}(1)$ value that we expect it to. We believe this is a purely technical problem\footnote{With better computing power (all our plots were generated on laptops) and more modern and less simple-minded collapse codes, we expect that this should be doable.} related to the code slowing down because of the smaller time steps needed at late times in these cases\footnote{Note that if the field range were always smaller than $\mathcal{O}(1)$, bulk EFT is trivially valid everywhere in the interior. What we are pointing out here is that this is an artefact, and that the scalar almost certainly diverges around the singularity, even in the numerically difficult cases. This is just the usual breakdown of bulk EFT at the singularity.}. In particular, we clearly see the correlated onset of divergence of the modulus and the Kretschmann scalar in every case\footnote{In fact, the localization of the growth of these quantities is often directly related to the difficulty in evolving the system. This is plausible because steep derivatives require tighter error management.}. So it is natural to expect that the growth will continue if our numerical tools were powerful enough to run for longer. 

\begin{figure}[H]
\begin{subfigure}{.5\textwidth}
  \centering
  \includegraphics[width=\linewidth]{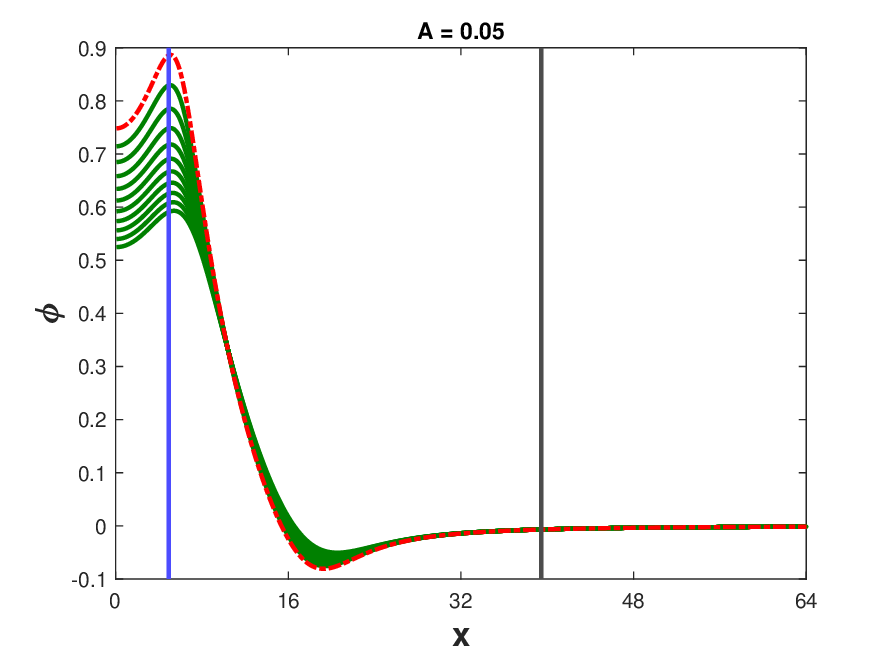}  
  \caption{$\phi$ vs $x$ for $A = 0.05$}
  \label{fig:Scalar_Vs_x_A_p05}
\end{subfigure}
\begin{subfigure}{.5\textwidth}
  \centering
  \includegraphics[width=\linewidth]{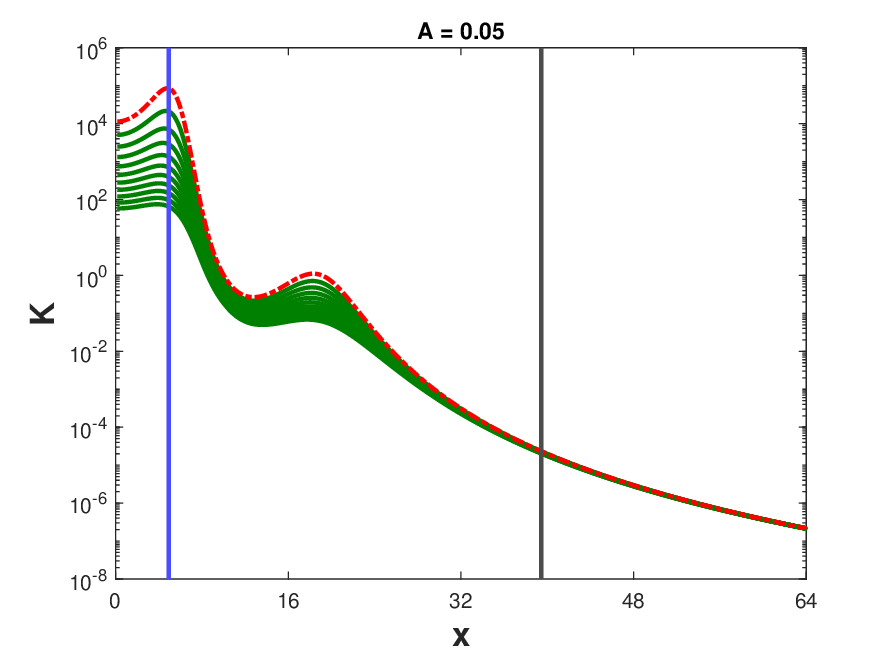}  
  \caption{$K$ vs $x$ for $A = 0.05$}
  \label{fig:K_Scalar_Vs_x_A_p05}
\end{subfigure}
\caption{Plot of modulus $\phi$ and Kretschmann scalar $K$ for $A = 0.05$, $\delta^{2} = 50$, $x_{0} = 100$ case, which we refer to as the Canonical Example. The $O(1)$ movement of the scalar is evident. The plots show the last few time steps (green curves) and the final time-step $t_{f}$ (the red dashed curve) at which the constraint conditions are satisfied within $\mathcal{O} \sim 10^{-3}$ (See Appendix \ref{Appendix-Constraints}). The numerical value of $t_f$ and the precise cut-off in error at which we stop the code do not affect our claims, so we do not quote those values. In all our figures blue and black vertical lines indicate the $x$ value of the peak of the Kretschmann scalar and the location of the outer apparent horizon, respectively, at $t_f$.}
\label{fig:phi_K_Vs_x_A_p_05}
\end{figure}

In some cases, we have noticed that the scalar peak/divergence and the Kretschmann divergence are not precisely coincident. But in these cases also, the apparent horizon is hierarchically separated from both.

\section{Concluding Comments}

The firewall paradox \cite{AMPS} is usually arrived at by starting with a smooth horizon in bulk effective field theory, and eventually finding a contradiction with one or more cherished principles like unitarity or locality. This leads to the conclusion that black holes older than the Page time have firewalls. The fuzzball program starts from the other end, and claims that black hole microstates have ``no interiors" to begin with. It is difficult to make a precise statement out of this in the full quantum (finite-$N$) theory in the bulk\footnote{The ``no interior" statement makes sense in the context of the supergravity states, but it is less clear what it might mean when the state is quantum. This is one reason why we have put quotes around ``no interior", in the abstract.}. So in \cite{Burman}, a Planckian stretched horizon was postulated as a UV regulator that captures this ignorance. Remarkably, it was found that the natural correlators on a microstate in this setting are effectively indistinguishable from the smooth horizon correlator in a bulk EFT until the Page time. One of our goals in this paper was to find an alternate argument with origins in the UV, for the smoothness of the horizons of young black holes in bulk EFT.

The usual argument for smoothness in bulk EFT has its roots in the principle of equivalence. But when we are trying to find the limits of applicability of this principle, we need to go beyond it. We looked for an idea which lets us make comments about EFTs based on the constraints arising from UV completion. Such an idea is the swampland distance bound \cite{Ooguri}. There is quite strong evidence that the swampland distance bound holds quite generally, for moduli scalars. To exploit this principle we studied the gravitational collapse of such moduli using the numerical methods suited for studying Choptuik-like systems \cite{CK}. We found that the region of  steep modulus growth during its evolution is hierarchically closer to the singularity and not the apparent horizon. This is an indication that bulk EFT is intact until we get close to the singularity. Put differently, if we were to look at the late-time curves in the scalar-gravity system, and use the swampland distance bound as a criterion for diagnosing the acceptability of these solutions, we would only remove the regions close to the singularity. 

We conclude with a technical comment. The discussions of \cite{Burman} were in the context of large black holes in AdS which are dual to heavy typical states in the CFT. These black holes are ``eternal" in the sense that they do not evaporate away and they have a time-translation symmetry in the exterior. When computing scalar correlators, the zero of the clock is therefore set by the insertion of the first scalar operator. This perturbation makes the state atypical, and the observation \cite{Burman} of the smooth horizon (a scrambling time after the initial insertion) is consistent with the claim of \cite{Susskind-typical} that horizons with increasing complexity are smooth. The situation we consider in this paper can be viewed as another realization of atypicality -- we are starting with a collapsing state which is by construction atypical. 

\section{Acknowledgments}

We thank Himanshu Chaudhary and Jun-Qi Guo for a previous related collaboration and help with the code, and Samir Mathur for encouraging comments on a previous draft.



\newpage
\appendix
\section{More Examples of $\boldsymbol{\mathcal{O}(1)}$ Movement}

We have done a fairly thorough scan of the Gaussian profile cases, and a less thorough scan of some other profiles. In all cases we find the same qualitative late-time behavior modulo the minor qualifications alluded to elsewhere. We plot some cases below to give the reader a sense of the evolutions.  

\begin{figure}[H]
\begin{subfigure}{.5\textwidth}
  \centering
  \includegraphics[width=\linewidth]{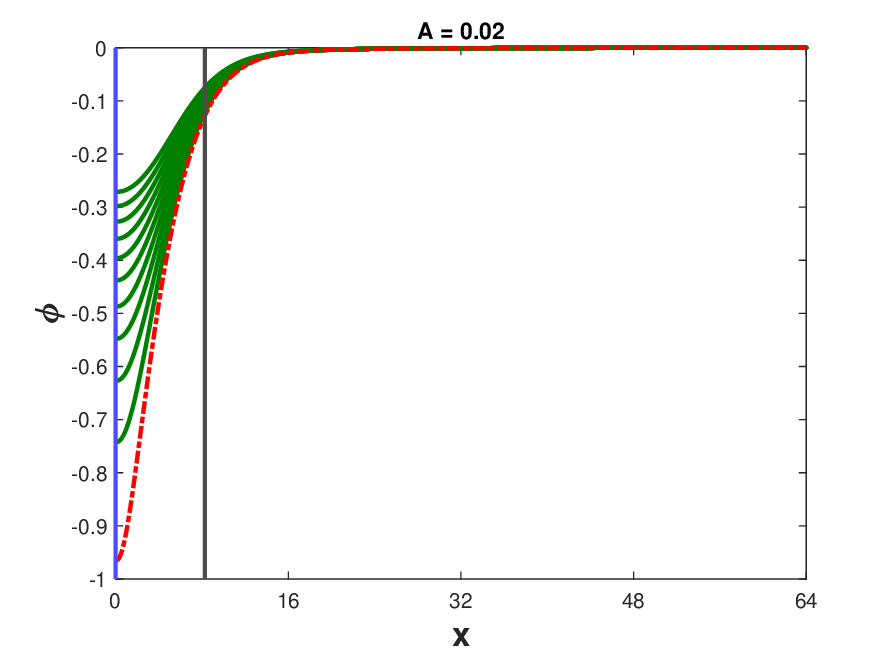}  
  \caption{$\phi$ vs $x$ for $A = 0.02$}
  \label{fig:psi_sum_vs_x_sum_A_p02}
\end{subfigure}
\begin{subfigure}{.5\textwidth}
  \centering
  \includegraphics[width=\linewidth]{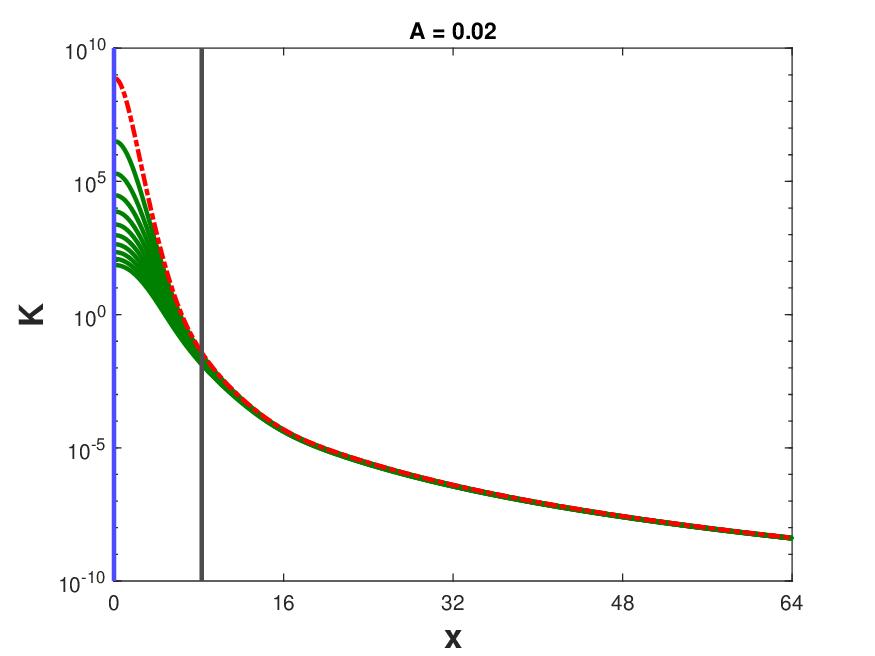}  
  \caption{$K$ vs $x$ for $A = 0.02$}
  \label{fig:K_sum_vs_x_A_p02}
\end{subfigure}
\newline
\begin{subfigure}{.5\textwidth}
  \centering
  \includegraphics[width=\linewidth]{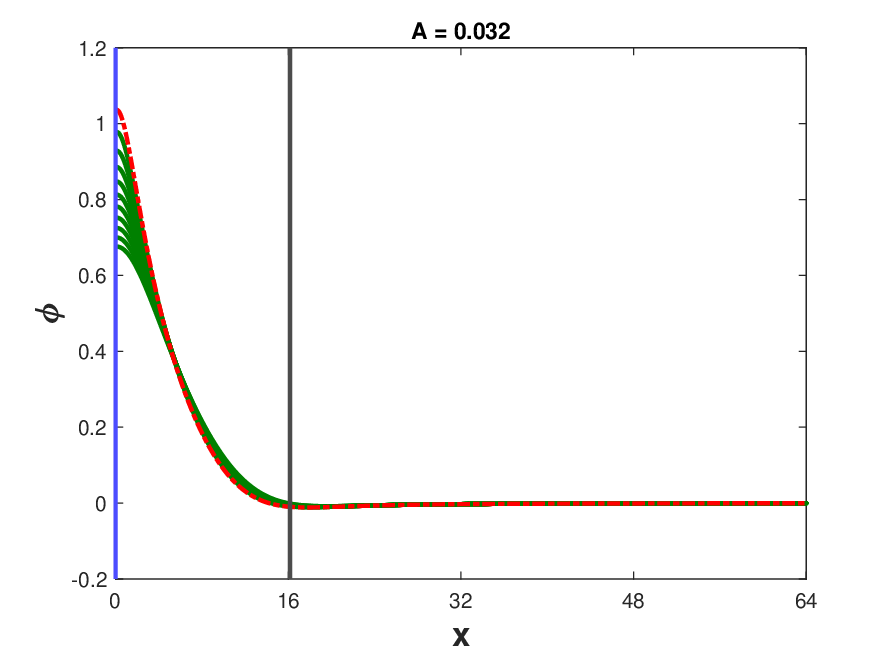}  
  \caption{$\phi$ vs $x$ for $A = 0.032$}
  \label{fig:psi_sum_vs_x_A_p032}
\end{subfigure}
\begin{subfigure}{.5\textwidth}
  \centering
  \includegraphics[width=\linewidth]{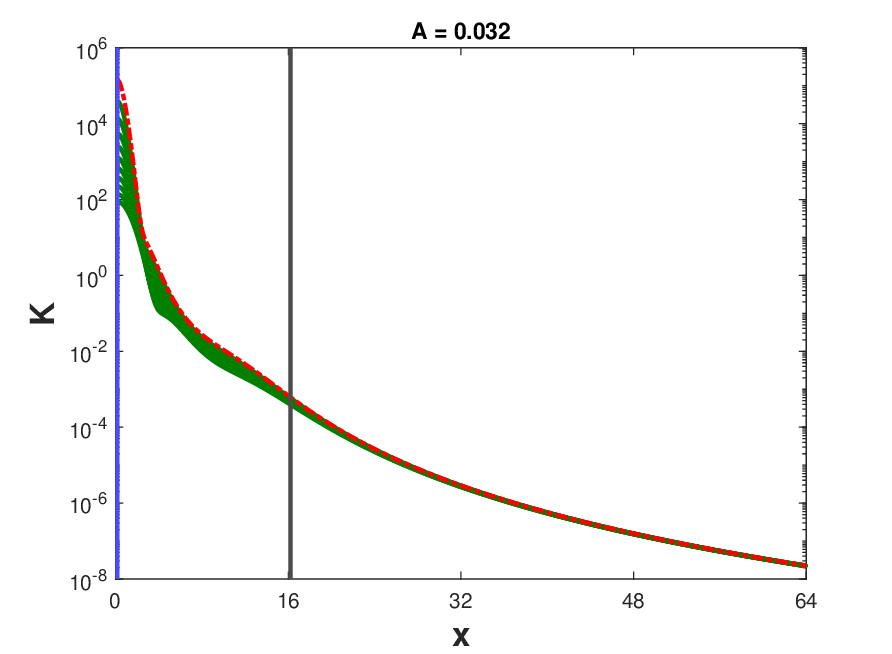}  
  \caption{$K$ vs $x$ for $A = 0.032$}
  \label{fig:K_sum_vs_x_sum_A_p032}
\end{subfigure}
\caption{Two examples of $\mathcal{O}(1)$ movement for $x_{0} = 100$, $\delta^{2} = 50$. The critical value in this case is $A=A_c\sim 0.016$, and the plots show that the scale separation is better for the more super-critical value of $A$. This observation is universally true, but the trade off is that (numerically) we find is easier to evolve the system when we are closer to criticality. Fortunately the curvature and the scalar movement are generally tightly correlated.}
\label{fig:phi_K_Vs_x_for_Other_O(1)_1}
\end{figure}

\begin{figure}[H]
\begin{subfigure}{.53\textwidth}
  \centering
  \includegraphics[width=\linewidth]{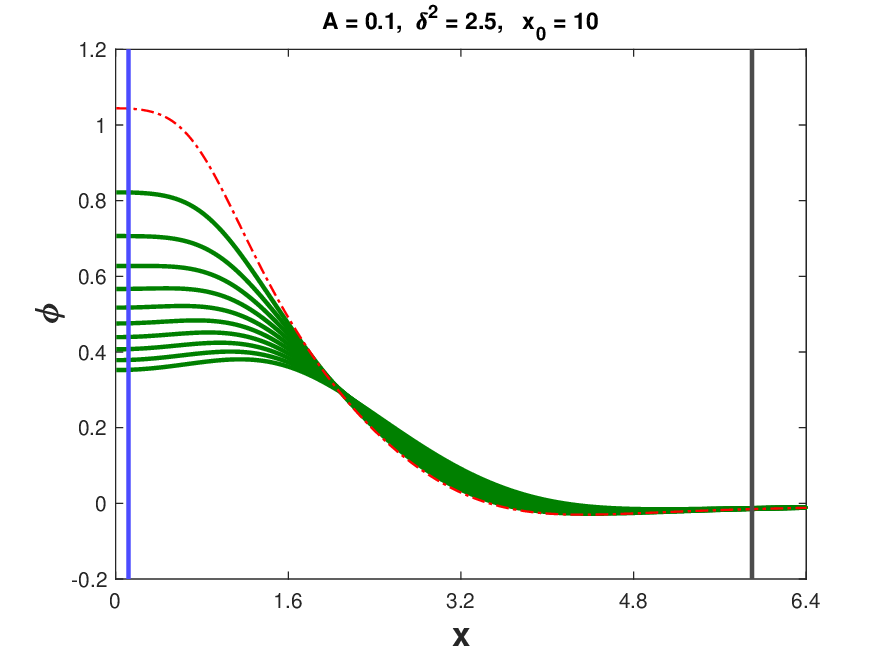}  
  \caption{}
  \label{fig:psi_sum_vs_x_sum_A_p02}
\end{subfigure}
\begin{subfigure}{.53\textwidth}
  \centering
  \includegraphics[width=\linewidth]{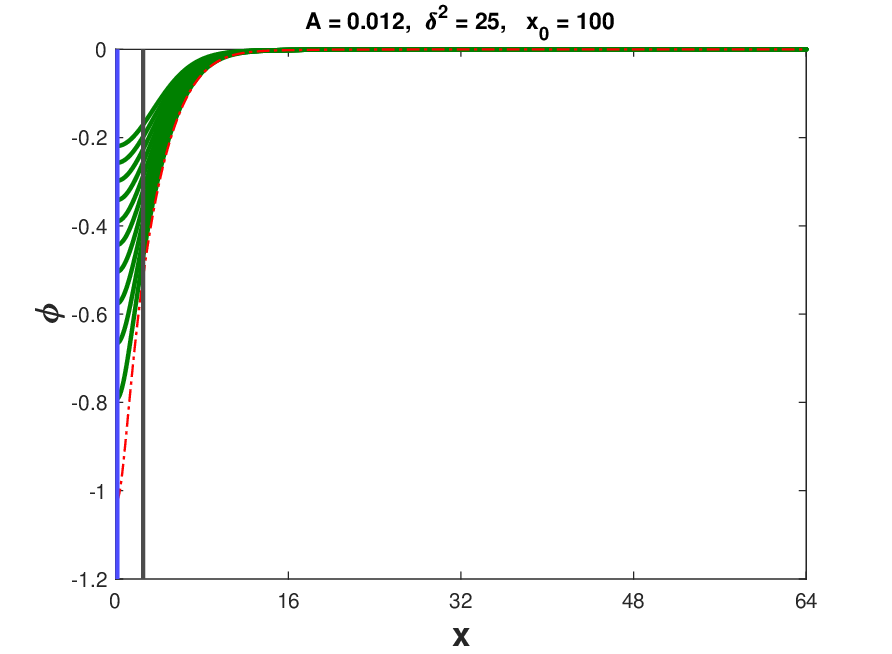}  
  \caption{}
  \label{fig:K_sum_vs_r_sum_A_p02}
\end{subfigure}
\newline
\begin{subfigure}{.53\textwidth}
  \centering
  \includegraphics[width=\linewidth]{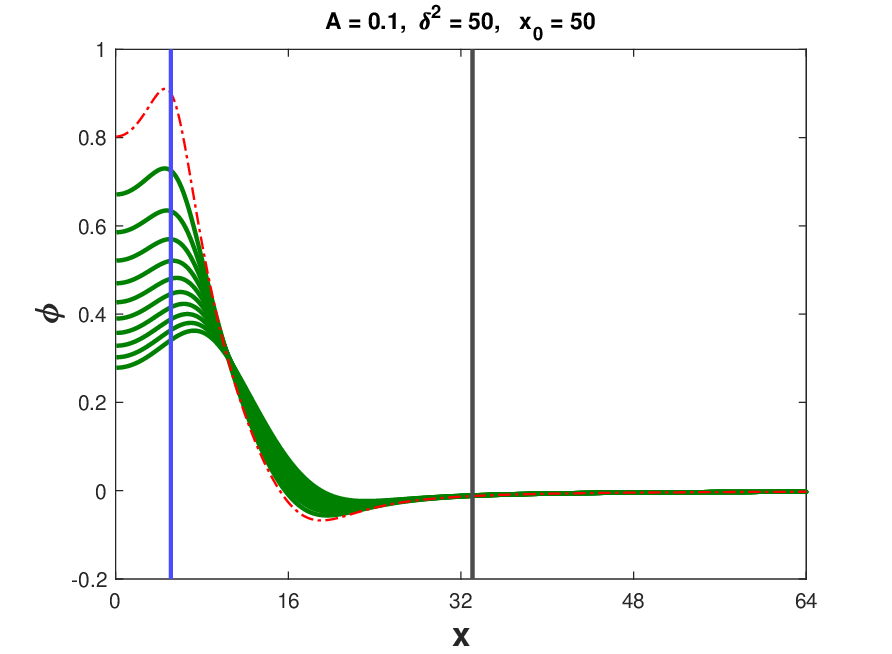}  
  \caption{}
  \label{fig:psi_sum_vs_x_sum_A_p032}
\end{subfigure}
\begin{subfigure}{.53\textwidth}
  \centering
  \includegraphics[width=\linewidth]{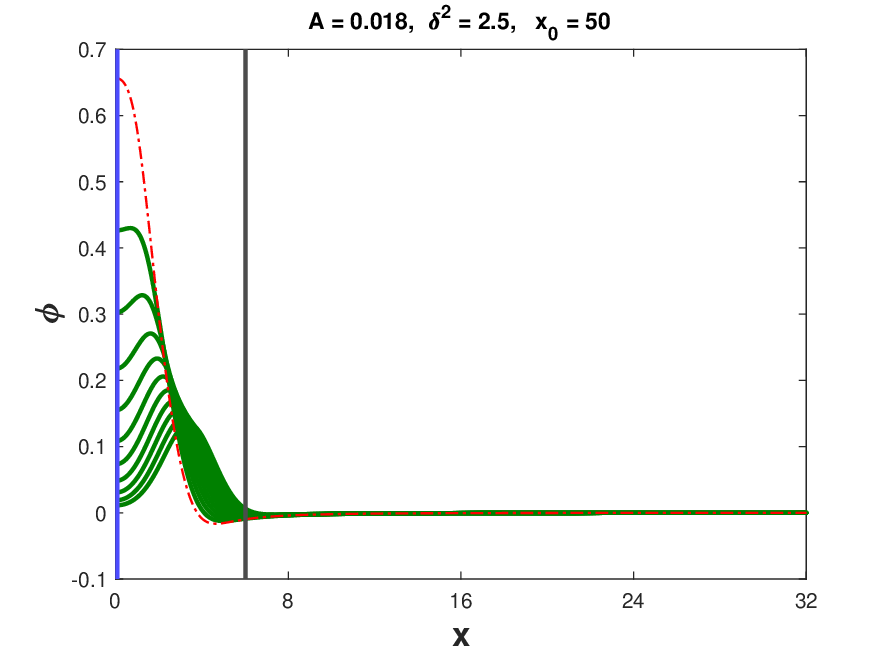}  
  \caption{}
  \label{fig:K_sum_vs_x_sum_A_p032}
\end{subfigure}
\caption{Some universality classes of late time curves.}
\label{fig:phi_K_Vsx_for_Other_O(1)_1}
\end{figure}

As mentioned in the main text, the final curves that we find in our evolutions fall into some ``universality classes" of shapes. In other words, many initial profiles lead to loosely similar final shapes. The figures in \ref{fig:phi_K_Vsx_for_Other_O(1)_1} are selected to indicate some of these final shapes that we see in our evolutions. While it is not important for our statements here, this universality may be of broader interest. ``Experimentally", we have found that some forms are significantly harder to evolve at late times than others -- in other words, in order to evolve by the same scalar range while respecting the constraints, we need much finer timesteps in some cases.


\section{Physical Measures of Distance}\label{PhysDis}

In this appendix, we give examples to illustrate that the statement about the separation of scales between scalar movement and the apparent horizon is not a coordinate artefact arising only in the $x$-coordinate. The most natural way to illustrate this is to work with the radial variable $r$ which defines the size of the sphere and is often used as a more physical choice of distance. We present the plots of the modulus and the Kretschmann scalar against $r$ below. It should be clear that the statements made earlier are still valid. 

\begin{figure}[H]
\begin{subfigure}{.5\textwidth}
  \centering
  \includegraphics[width=\linewidth]{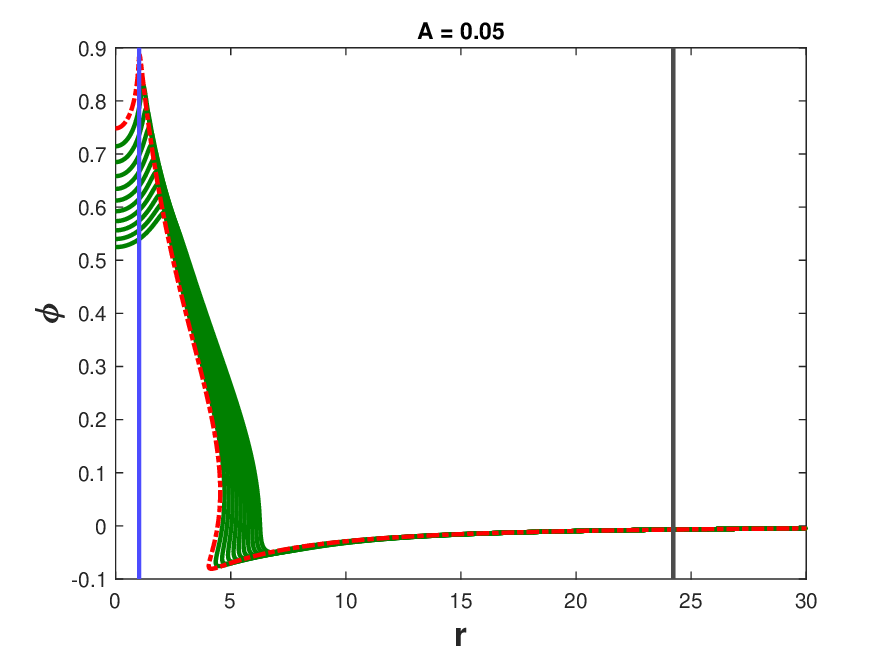}  
  \caption{$\phi$ vs $r$ for $A = 0.05$}
  \label{fig:Scalar_Vs_r_A_p05}
\end{subfigure}
\begin{subfigure}{.5\textwidth}
  \centering
  \includegraphics[width=\linewidth]{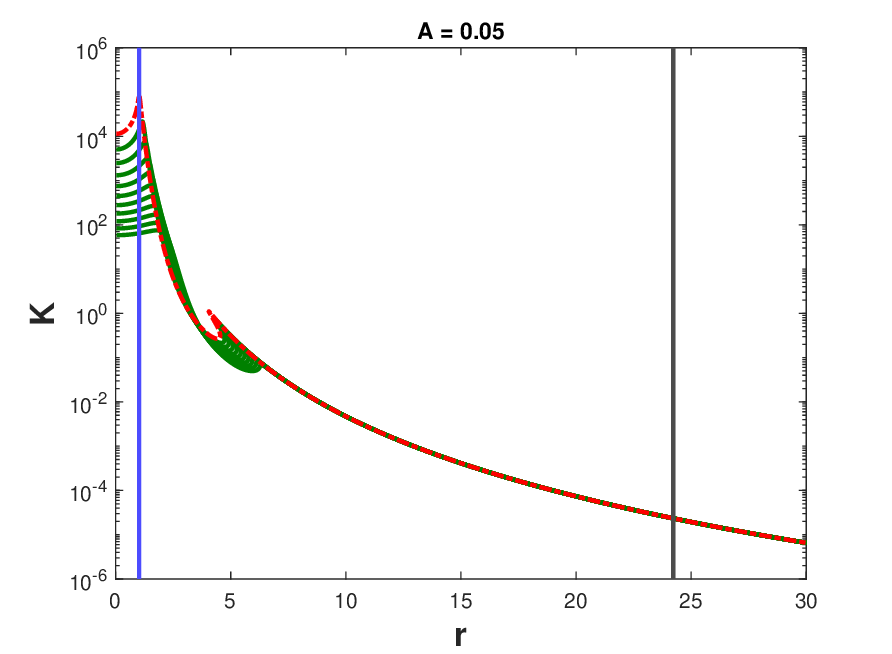}  
  \caption{$K$ vs $r$ for $A = 0.05$}
  \label{fig:K_Scalar_Vs_r_A_p05}
\end{subfigure}
\caption{Plot of the scalar field $\phi$ and the Kretschmann scalar $K$ as a function of the radial function $r(t,x)$ for the Canonical Example.}
\label{fig:phi_K_Vs_r_A_p_05}
\end{figure}
\begin{figure}[H]
    \centering
    \includegraphics[width=0.5\linewidth]{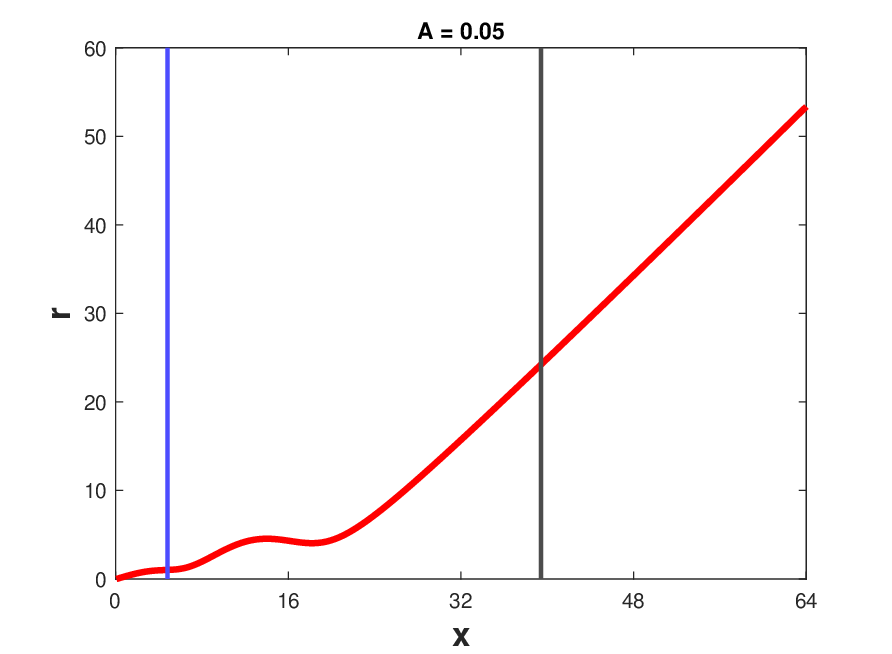}
    \caption{Plot of the radial function $r(t_{f},x)$ for $A = 0.05$, $\delta^{2} = 50$, $x_{0} = 100$ case. The vertical lines indicate the same features mentioned in Fig. \ref{fig:phi_K_Vs_x_A_p_05}. }
    \label{fig:r_Vs_x_p_05}
\end{figure}

The cusps and the turnarounds in the above figures may be disturbing (which is one reason why we have presented the $x$-plots in the main text). But they are physical. Note that the radial variable is not a coordinate, it is a dynamically determined metric function. In other words, it is a function of $(t, x)$ that one obtains by evolving the Einstein equations. This means that $r(t,x)$ need not be a monotonic function of $x$ for a given $t$. This is illustrated in Figure \ref{fig:r_Vs_x_p_05}.

Another natural measure of invariant distance is provided by the integral of the line element $ds$ itself, at fixed $t$ and angular coordinates. This is the quantity noted in \eqref{invdis}. It is easy enough to make plots against $s$ and check that the curves retain the qualitative features of the modulus growth and the location of the apparent horizon that we have mentioned earlier. But it is also easy to see it intuitively from the plot of $e^{-\sigma}$ below. It is evident that the area under the curve between the two relevant vertical lines is $\mathcal{O}(1)$ and therefore the $x$-range and $s$-range are both of the same order.
\begin{figure}[H]
    \centering
    \includegraphics[width=0.5\linewidth]{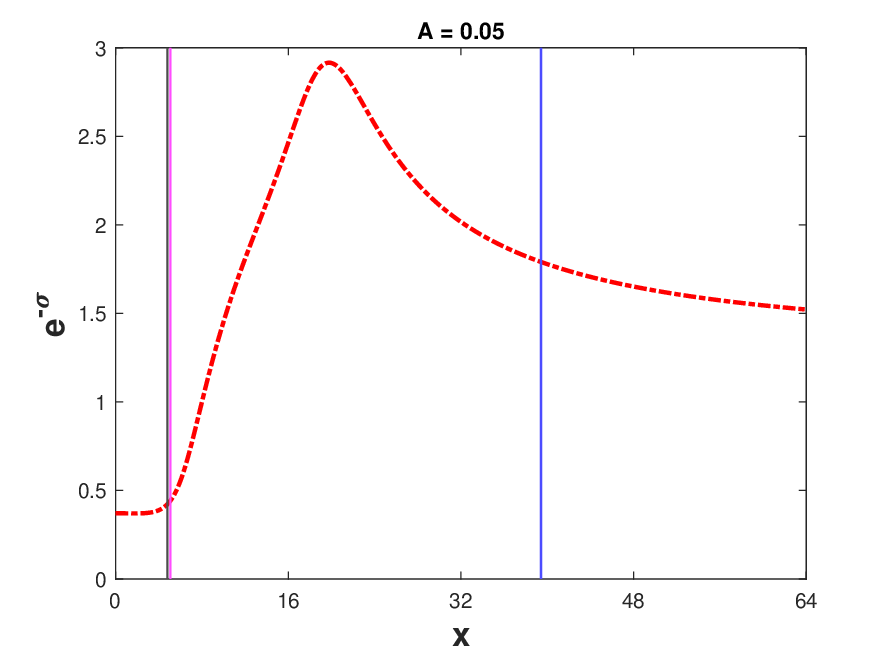}
    \caption{Plot of $\exp{(-\sigma(t_{f},x)})$ for the Canonical Example. $s = \int \exp{-\sigma(t_{f},x)} dx$ is the measure of separation between two $x$ coordinates. We found the separation between the singularity (black vertical line) and the $\mathcal{O}(1)$ peak of the scalar (magenta line) is $\sim 0.116$, whereas that between the singularity and the apparent horizon (blue vertical line) is $\sim 70$.}
    \label{fig:exp_m_sigma_sum_A_p_05}
\end{figure}


\section{Constraints}\label{Appendix-Constraints}
\begin{figure}[H]
\begin{subfigure}{.5\textwidth}
  \centering
  \includegraphics[width=\linewidth]{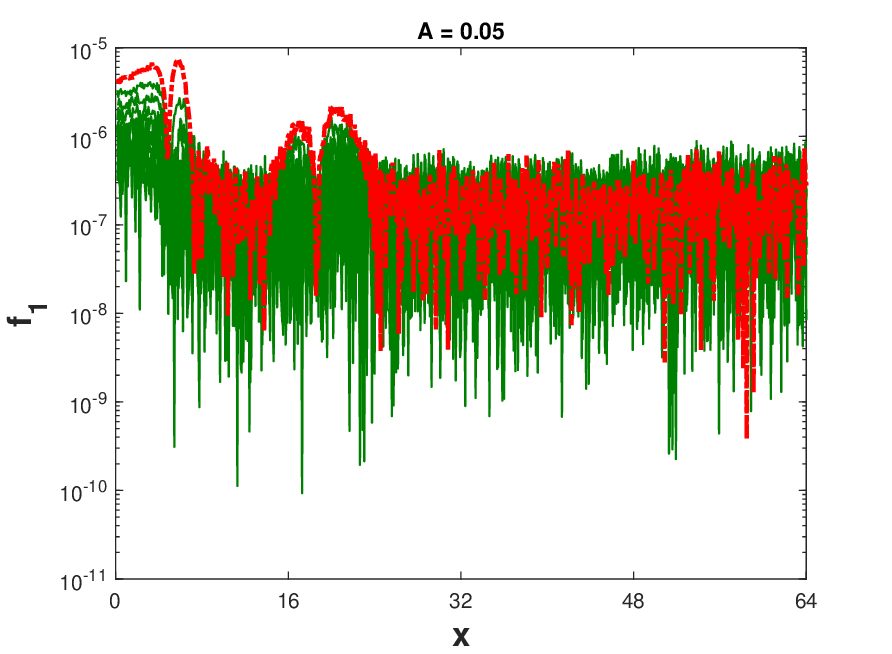}  
  \caption{$f_{1}$ vs $x$ for $A = 0.05$}
  \label{fig:Constraint_1_Vs_x_A_p05}
\end{subfigure}
\begin{subfigure}{.5\textwidth}
  \centering
  \includegraphics[width=\linewidth]{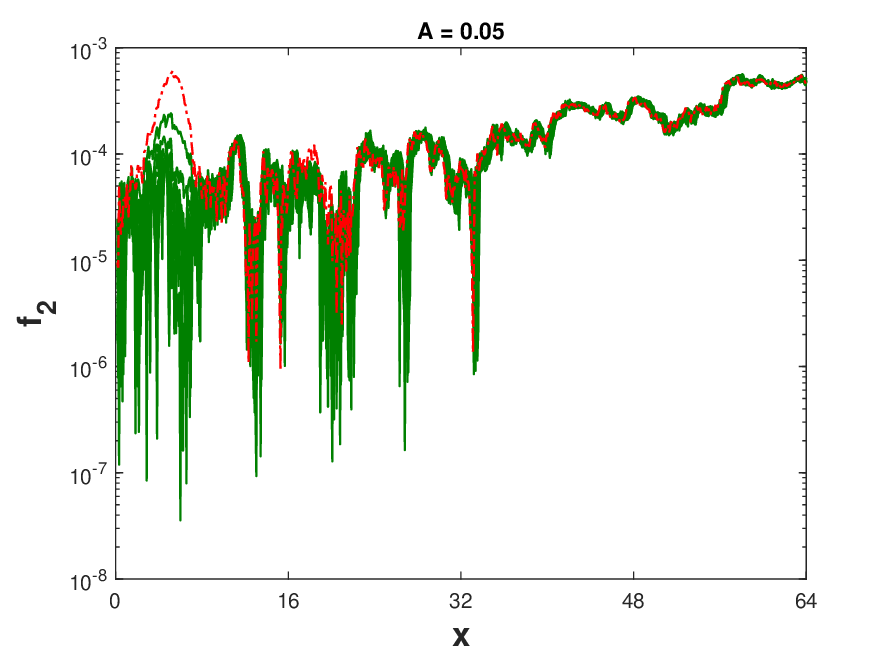}  \caption{$f_{2}$ vs $r$ for $A = 0.05$}
  \label{fig:Constraint_2_Vs_x_A_p05}
\end{subfigure}
\caption{Plot of the constraints $f_{1}$ and $f_{2}$ for time slices shown in Fig. \ref{fig:phi_K_Vs_r_A_p_05}. The constraints remain well satisfied within the $x$ range where the large field movement occurs. Red curve denotes the final time step as before.}
\label{fig:Constraint_Vs_x_A_p_05}
\end{figure}

We evaluate the two Einstein constrains after each time step. Because we are deep in the super-critical regime, in order to see the $\mathcal{O}(1)$ field growth, we have often had to go closer to the singularity than in \cite{CK}. When the evolution gets close to the singularity, the errors in the constraints start increasing due to the steep derivatives. To proceed, we must refine the time-steps in our code. Fortunately, since the swampland distance bound is a semi-qualitative statement, we can use the fairly simple-minded integration used in \cite{CK} here as well, but with extra care paid to the constraints and time-step management. But it should be emphasized that this is one of our biggest bottlenecks -- a more sophisticated approach, such as adaptive mesh refinement, will be required if one wants to make more refined statements.

\bibliographystyle{JHEP} 
\bibliography{references}

\end{document}